# Micron-scale D/H heterogeneity in chondrite matrices: a signature of the pristine solar system water?

*Laurette PIANI[*, 1], François ROBERT, Laurent REMUSAT*

Institut de Minéralogie, de Physique des Matériaux et de Cosmochimie, Sorbonne Universités, UMR CNRS 7590, Université Pierre et Marie Curie, IRD, Muséum National d'Histoire Naturelle, 57 rue Cuvier, Case 52, 75231 Paris Cedex 5, France.

[*]Corresponding author

[1] Present address: Department of Natural History Sciences, Hokkaido University, Sapporo 060-0810, Japan. E-mail address: laurette@ep.sci.hokudai.ac.jp (L. Piani)

**ABSTRACT**

Organic matter and hydrous silicates are intimately mixed in the matrix of chondrites and in-situ determination of their individual D/H ratios is therefore challenging. Nevertheless, the D/H ratio of each pure component in this mixture should yield a comprehensible signature of the origin and evolution of water and organic matter in our solar system.

We measured hydrogen isotope ratios of organic and hydrous silicates in the matrices of two carbonaceous chondrites (Orgueil CI1 and Renazzo CR2) and one unequilibrated ordinary chondrite (Semarkona, LL3.0). A novel protocol was adopted, involving NanoSIMS imaging of H isotopes of monoatomatic (H$^-$) and molecular (OH$^-$) secondary ions collected at the same location. This allowed the most enriched component with respect to D to be identified in the mixture. Using this protocol, we found that in carbonaceous chondrites the isotopically homogeneous hydrous silicates are mixed with D-rich organic matter. The opposite was observed in Semarkona. Hydrous silicates in Semarkona display highly heterogeneous D/H ratios, ranging from 150 to 1800 × 10$^{-6}$ (δD$_{SMOW}$ = -40 to 10,600 ‰). Organic matter in Semarkona does not show such large isotopic variations. This suggests limited isotopic exchange between the two phases during aqueous alteration. Our study greatly expands the range of water isotopic values measured so far in solar system objects. This D-rich water reservoir was sampled by the LL ordinary chondrite parent body and an estimate (≤9%) of its relative contribution to the D/H ratio of water in Oort cloud family comets is proposed.

## 1. Introduction

Water and/or organic molecules have been observed in pre-stellar clouds, protostars and protoplanetary disks (Herbst and van Dishoeck, 2009). The hydrogen isotopic composition of water and organic molecules is a powerful tool for deciphering the different steps in the evolution of matter throughout these primordial environments (e.g. Ceccarelli et al., 2014 and Bergin et al., 2006 and references therein). Fractionations due to isotopic exchange in the gas phase, between gas and grains, or at the surface of the grains, are thought to take place in pre-stellar clouds or during protostar and disk evolution, resulting in high deuterium enrichments in water and organics. Preserved in ices, these isotopic enrichments can be partly recovered in comets as well as in undifferentiated meteorites.

However, astrophysical and cosmochemical studies have shown that the distribution of the D/H ratio throughout solar system materials is complex. In the gas phase, the main H-bearing component, $H_2$, presents the lowest D/H value ($[20 \pm 3.5] \times 10^{-6}$; Geiss & Gloecker, 2003). The other H-carriers measured in the solar system so far are all enriched in deuterium compared to this value (for a review, see Robert, 2006). The bulk Earth D/H ratio (BE) is estimated to be $[149 \pm 3] \times 10^{-6}$ (Lecuyer et al., 1998), a value similar to the mean isotopic composition of bulk carbonaceous chondrites. However, higher D/H ratios have been locally measured in hydrous minerals in some chondrites, demonstrating that the water isotopic composition in chondrites is in fact heterogeneous (Deloule and Robert, 1995; Deloule et al., 1998; Grossman et al., 2000; Yurimoto et al., 2014; Bonal et al., 2013).

In comets, water has been found to be either similar to or enriched in deuterium compared to BE. Oort cloud family comets display a D-enrichment compared to BE (D/H ~ $300 \times 10^{-6}$, Bockelée-Morvan et al., 1998, 2012; Meier et al., 1998b). The Jupiter family comet Hartley 2, which may have acquired its ice in an outer environment, also exhibits a water D/H ratio close to BE (Hartogh et al., 2011). Several authors have therefore proposed that carbonaceous chondrites and comets might be considered members of the same "family" (e.g. Gounelle, 2011). This has been supported by mineralogical and isotopic analyses of dust grains returned to Earth by the Stardust space mission (McKeegan et al., 2006).

On Earth, organic molecules in isotopic equilibrium with water exhibit systematically lower D/H ratios relative to water. The reverse is observed in extraterrestrial environments (Ceccarelli et al., 2014). For example, HCN in the Hale-Bopp comet has a D/H ratio of $[2300 \pm 0.4] \times 10^{-6}$ (Meier et al., 1998a) compared to $300 \times 10^{-6}$ in water.

In carbonaceous chondrites, the main organic component is the Insoluble Organic Matter (IOM), which can be isolated from the matrix by acid treatment. This IOM is generally enriched in deuterium compared to water (Alexander et al. 2012). In CI and CR carbonaceous chondrites, the isotopic composition of the IOM is heterogeneous at the micron scale and "hotspots" have been reported with D/H ratios of up to $3600 \times 10^{-6}$ (Busemann et al., 2006; Remusat et al., 2009). Such large enrichments have also been reported in organic matter in Interplanetary Dust Particles (IDPs) (up to $4020 \times 10^{-6}$) and Ultra-Carbonaceous MicroMeteorites (UCAMMs) (up to $4600 \times 10^{-6}$) (Aléon et al., 2001; Duprat et al., 2010; Messenger, 2000).

In the past two decades, several models have attempted to describe the heliocentric evolution of the D/H ratio as a function of time in the disk. The aim of these models was to compare cometary and chondritic data with theoretical predictions (e.g. Lécluse and Robert, 1994; Aikawa and Herbst, 1999; Drouart et al., 1999; Mousis, 2000; Hersant et al., 2001; Willacy, 2007; Horner et al., 2007; Willacy and Woods, 2009; Jacquet and Robert, 2013; Yang et al., 2013, Cleeves et al. 2014). The major differences between these models concern (i) the structure and lifetime of the disk and (ii) the initial solar system water D/H ratio (hereafter referred to as $(D/H)_{t=0}$). Because the isotopic enrichment of water in contact with molecular $H_2$ is progressively lowered in the inner disk (i.e. $H_2 + HDO \rightarrow H_2O + HD$), all of these models postulate that $(D/H)_{t=0}$ was much higher than the protosolar value, reflecting possible heritage of ion-molecule reactions in the parent molecular cloud (D/H ~ $10^{-2}$ in a low mass protostar) or at the surface of the disk (estimated to be up to $6.4 \times 10^{-2}$, see review by Bergin et al. 2006). Based on a D/H ratio obtained by ion probe on H-bearing silicates in the LL3 Semarkona meteorite (Deloule and Robert, 1995), several models adopted $(D/H)_{t=0} \approx 10^{-3}$ as a reference value for their numerical simulations (Drouart et al., 1999; Hersant et al., 2001; Mousis, 2000; Yang et al., 2013) or for discussions of their theoretical results (Aikawa and Herbst, 1999; Willacy and Woods, 2009; Willacy, 2007). However, taking the D/H ratio of Semarkona phyllosilicates as a face value for initial solar system water implies that these phyllosilicates were formed in contact with primitive water and avoided D/H resetting during parent body evolution. The nature of the interactions between hydrogen-bearing species, such as ice, hydrated minerals or organics, is therefore addressed in the present paper.

It is commonly accepted that the decay of short-lived radionuclides (mostly $^{26}$Al) provided an important heat source, resulting in the melting of ice accreted with dust on the asteroid parent-body (Hutcheon and Hutchison, 1989). However, the extent of the resulting aqueous alteration is a matter of debate. As carefully detailed in Bland et al. (2009), geochemical and petrographic observations of unequilibrated chondrites argue for isochemical alteration occurring in a closed-system and at a restricted scale (≤ a few 100 μm) (McSween, 1979, Brearley, 2003), while numerical models indicate that alteration was open-system with liquid transport over tens of kilometers (e.g. Young et al., 1999). With regard to the hydrogen isotopic composition, the extent of aqueous alteration is important because large-scale flows of liquid would rapidly homogenize the isotopic composition and favor isotopic exchange between H-bearing phases such as phyllosilicates and organic matter.

Because of their intimate mixing at the nanometer-scale (Le Guillou and Brearley, 2013; Le Guillou et al., 2014), direct determination of individual isotope compositions of organics and hydrated minerals is challenging (e.g. Remusat et al., 2010; Alexander et al., 2010, 2012; Bonal et al., 2013; Le Guillou and Brearley, 2013), even using the highest spatial resolution techniques (Remusat et al. 2010; Bonal et al., 2013).

In the present study, we therefore used a protocol recently developed with the NanoSIMS (Piani et al., 2012) to identify the most deuterium-rich component in the matrices of two carbonaceous chondrites, Orgueil and Renazzo, and one unequilibrated ordinary chondrite, Semarkona. These primitive meteorites represent different degrees of aqueous alteration and their H-bearing silicates and organics have been shown to cover large ranges of hydrogen isotopic compositions (e.g. Alexander et al., 2010; Remusat et al., 2009; Robert, 2006).

We address two critical pending issues: Are the isotopic variations observed at a micron-scale resolution caused by variations in the organic/hydrous silicate ratio or by variations in their respective D/H ratios? And are these isotopic variations compatible with a model of large-scale liquid water flow on the asteroidal parent-body? Finally, we discuss how our results can be used to constrain the initial distribution of D/H ratios in the early solar system.

## 2. Experimental methods

2.1. Samples

Three unequilibrated chondrites were selected for this study: Orgueil (CI1), Renazzo (CR2) and Semarkona (LL3.0). CI carbonaceous chondrites are the most hydrated chondrite group (Brearley, 2006). Orgueil contains up to 15 wt. % water (Jarosewich, 1990). In this meteorite, the organic matter appears to have been partly redistributed by fluids as a diffuse organic component (Le Guillou et al., 2014). Aqueous alteration occurred on the CR parent body with variable degrees of alteration (Weisberg et al., 1993). In Renazzo, the organic matter is mostly found in the form of individual grains, attesting to its low mobilization by fluids (Le Guillou et al., 2014). Clay minerals such as serpentine and smectite are the main alteration products of carbonaceous chondrite matrices (Barber, 1981; Brearley, 2006). Fe-rich smectite is observed in a few ordinary chondrites such as Semarkona or Bishunpur (LL3.15) as the main H-bearing alteration product of fine-grained silicates and sulfides in the matrix, and to a lesser extent in the mesostasis of a few chondrules (Grossman et al., 2000; Hutchison et al., 1987). The Semarkona matrix also contains poorly-organized organic matter (Quirico et al., 2003, Le Guillou et al. 2012).

Pieces of matrix of a few hundred microns in size were handpicked using a binocular microscope and pressed in pure gold or indium foil. Using a polarized reflected-light microscope, zones of fine-grained matrix were selected in order to avoid large minerals or chondrule fragments and imaged with the NanoSIMS (one large zone in Orgueil and Renazzo and three in Semarkona). After NanoSIMS measurement, the three zones of Semarkona were observed using a scanning electron microscope in order to verify the fine-grained nature of the imaged areas and the presence of elements characteristic of smectite, such as silicon, iron and sodium. Because of the fine-grained nature of chondrite matrix, it was not possible to precisely identify the composition of the mineral phases imaged by NanoSIMS. However, due to their ubiquity in aqueous altered matrices and their high hydrogen content, phyllosilicates (serpentine and smectite) can be considered the main contributors of inorganic hydrogen.

A type III kerogen (hereafter referred to as Ker) and the IOM of the Antarctic Grosvenor Mountains 95502 ordinary chondrite (IOM-Gro) were used as standards for the organic matter component (Piani et al., 2012). A montmorillonite smectite (Mtm) from Camps Bertaux, Morocco, was chosen as the reference phyllosilicate component. A laboratory-prepared mixture of 10% IOM-Gro and 90% Mtm (IOM10) was also analyzed during the same analytical sessions.

2.2. Analytical protocol

We applied the protocol detailed in Piani et al. (2012), which is based on the difference in intensity of the $H^-$ and $OH^-$ ion species coming from H-bearing silicates and organic matter under a $Cs^+$ primary beam ($H^-/^{16}OH^-$ of 1.6 and 12.5 in montmorillonite and organic matter, respectively, from Table 2 of Piani et al., 2012). Hence, the measured $D^-/H^-$ and $^{16}OD^-/^{16}OH^-$ ratios vary differently with the phase proportions and isotopic compositions of organic matter and H-bearing silicates in a mixture of both phases. A schematic view of the protocol is presented in Figure 1. The protocol was tested using laboratory-prepared mixtures of Mtm and IOM-Gro and was shown to be suitable for estimating the D/H composition of the two intermixed components.

NanoSIMS images were obtained on a mixture of organic matter and phyllosilicates as well as on pure natural organic matter and phyllosilicates (i.e., samples mainly composed of either organic matter or phyllosilicates). $H^-$ and $^{16}OH^-$ intensities and $D^-/H^-$ and $^{16}OD^-/^{16}OH^-$ ratios were measured for all samples and standards under identical analytical conditions. Representation of the results in a $D^-/H^-$ ratio versus $^{16}OD^-/^{16}OH^-$ ratio plot allows rapid qualitative information to be obtained (Figure 1). Measured $D^-/H^-$ and $^{16}OD^-/^{16}OH^-$ ratios form either (i) a linear trend for pure components or (ii) a mixing curve for mixtures due to the difference in $H^-$ and $^{16}OH^-$ emissivity from organic matter and phyllosilicates. According to the position of the mixing curve relative to the linear correlation of pure components, it is possible to determine whether the highest D/H ratio is hosted by organic matter or by hydrous minerals.

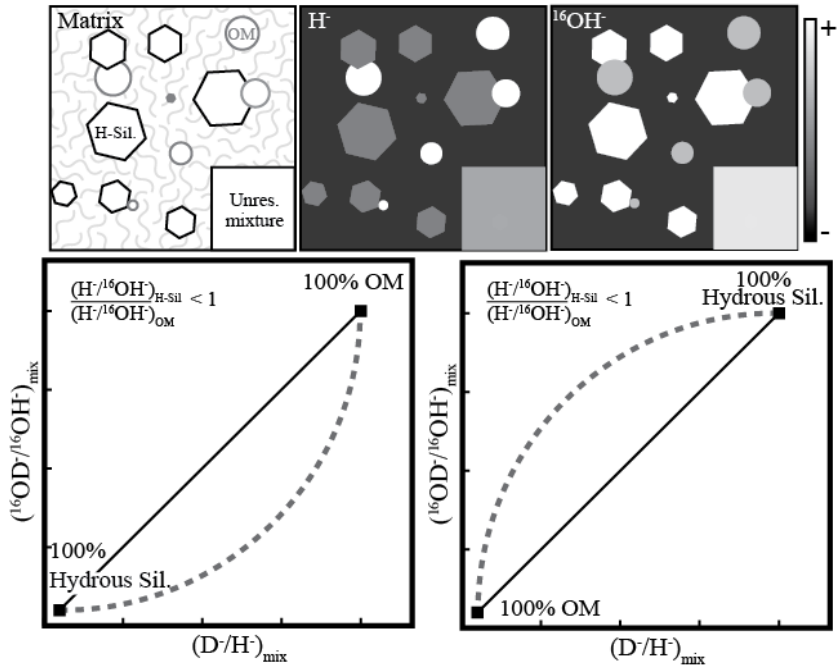

Figure 1. Schematic view of the protocol detailed in Piani et al., (2012) and used in this study. Upper part: cartoon representing a matrix with organic matter (OM, circle symbols), hydrous silicates (H-Sil, polygons) and an area where the mixture cannot be resolved by the NanoSIMS (Unres. mixture, large square). The intensity of $H^-$ and $^{16}OH^-$ ions when imaged using a $Cs^+$ primary beam are shown in two illustrations on the right. Lower part: representation of the measured $D^-/H^-$ and $^{16}OD^-/^{16}OH^-$ ratios in a mixture with proportions from 0 to 100% of each phase. On the left, the organic matter is the D-rich end-member and, on the right, hydrous silicates are the D-rich end-member.

2.3. NanoSIMS analytical conditions and image processing

Ion images were obtained with the NanoSIMS 50 installed at MNHN in Paris. Prior to analysis, five minutes of high primary current (500 pA) presputtering was applied in order to clean the surface and reach the sputtering steady-state. Following this, 10 × 10 μm² images were acquired in two steps with a 16 keV $Cs^+$ ion primary beam. Images of 128 × 128 pixels were obtained with a 3-ms counting time per cycle and per pixel. The electron-flooding gun was used to improve charge compensation. In the first step, $^{13}C^-$, $^{16}OH^-$, $^{16}OD^-$, and $^{28}Si^-$ secondary ions were collected with a 65-70 pA primary current resulting in a beam diameter of ~ 500 nm and high mass resolution settings ($MRP_{cam}$ = 12,000, where $MRP_{cam}$ is the Cameca's definition of the MRP, see for example Fletcher et al., 2008). This high mass resolution allows the $^{17}OH^-$ interference on $^{16}OD^-$ to be reduced to less than 5% of $^{16}OD^-$ intensity, resulting in a maximum relative uncertainty of 0.0082 in the $^{16}OD^-/^{16}OH^-$ ratio. In the second step, $H^-$ and $D^-$ were collected in the same areas with a 20 pA primary current, a beam diameter of ~ 400 nm and a mass resolution $MRP_{cam}$ = 5000. The total acquisition duration for the two measurement steps on a single 10 × 10 μm² area is around 1.5 hours.

To measure the carbonaceous chondrites and the ordinary chondrite, two sets of analyses were performed at different times but with similar settings. Each time, IOM-Gro, Mtm, IOM10 and Ker were measured as reference samples for intensity and isotopic ratios.

Images were processed using the L'IMAGE software package (Larry Nittler, Washington). Data were extracted from 4 × 4 μm² regions of interest (ROI) defined in the center of the images, resulting in four ROIs per image. Reference sample data are given in supplementary online material Table S1. Uncertainty for each ROI was calculated at one sigma, taking into account the statistical error, the reproducibility calculated on reference samples and the uncertainty due to $^{17}OH^-$ interference for the $^{16}OD^-/^{16}OH^-$ ratios. In addition, for Semarkona, ROIs were also defined around isotopically heterogeneous areas (Figure S1).

**3. Results**

3.1. Matrix of Orgueil and Renazzo carbonaceous chondrites

The $H^-$, $^{13}C^-$, $^{16}OH^-$ and $^{28}Si^-$ distributions and corresponding $D^-/H^-$ and $^{16}OD^-/^{16}OH^-$ ratio images are shown in Figure 2 (data in Table S2). In the Orgueil and Renazzo NanoSIMS images, we observe the ubiquitous presence of silicon while carbon is randomly distributed as micrometer-scale spots. The distributions of $H^-$ and $^{16}OH^-$ do not appear to be correlated and match neither the $^{13}C^-$ nor $^{28}Si^-$ distributions. The isotopic images show a homogeneous distribution for the $^{16}OD^-/^{16}OH^-$ ratio, while micrometer-scale heterogeneities can be observed in the $D^-/H^-$ images. The lowest $D^-/H^-$ and $^{16}OD^-/^{16}OH^-$ ratios were obtained in Orgueil. For each carbonaceous chondrite, the deuterium content increases linearly with the organic matter content of the ROI, as illustrated in Figure 3.A. For comparison, data obtained for the reference samples IOM-Gro, Mtm and IOM10 are also reported. Orgueil shows a continuous increase from the terrestrial Mtm D/H ratio through the IOM10 mixture and towards D-rich chondritic IOM, while the Renazzo trend is displaced towards values significantly enriched in deuterium.

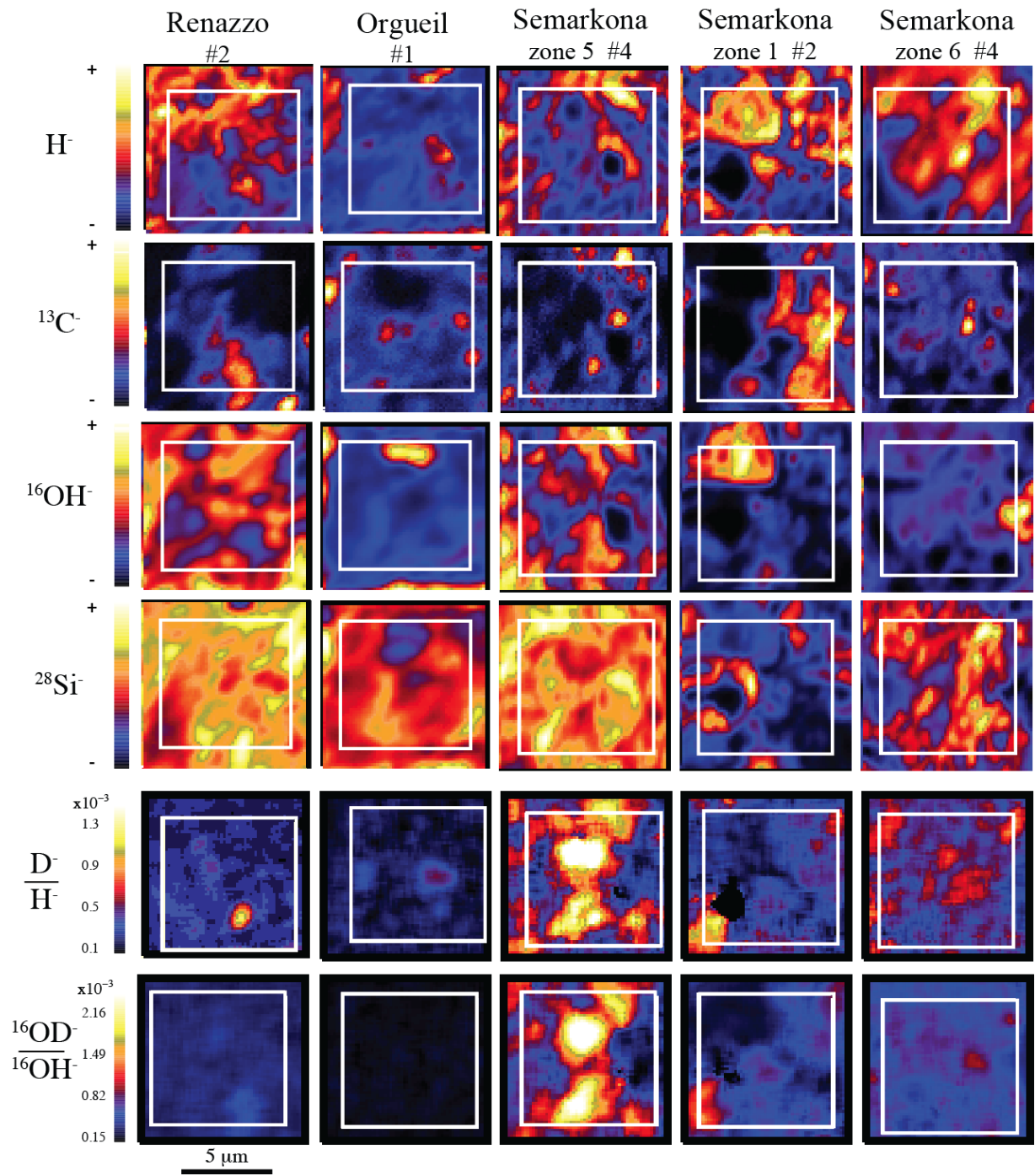

Figure 2. NanoSIMS images of Orgueil, Renazzo and Semarkona (for three different matrix zones): H$^-$, $^{13}$C$^-$, $^{16}$OH$^-$ and $^{28}$Si$^-$ distributions (arbitrary scale) and the corresponding D$^-$/H$^-$ and $^{16}$OD$^-$/$^{16}$OH$^-$ ratio images (identical scale for all meteorites). The image number (#) of each sample is indicated for reference to the data in Table S2.

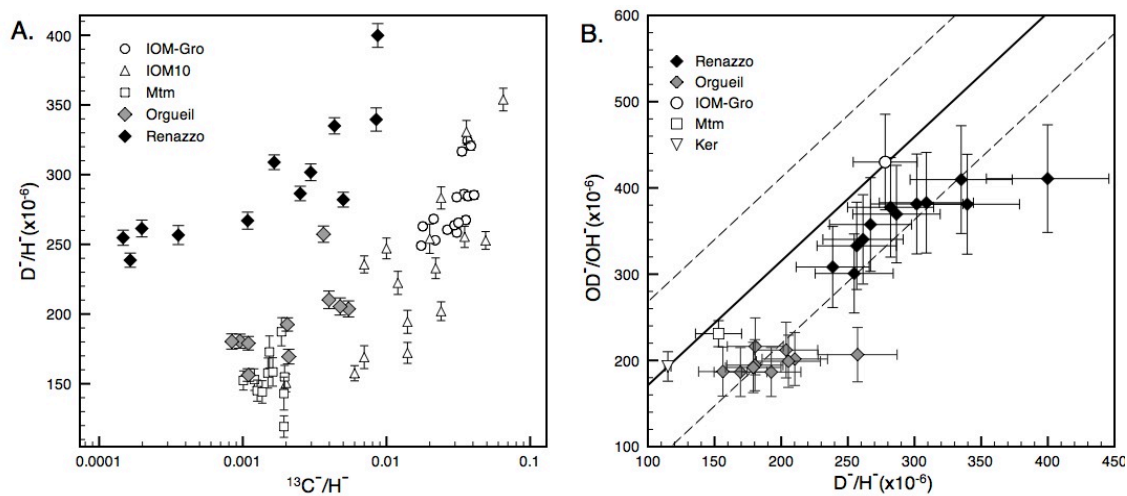

Figure 3. (A) Measured D⁻/H⁻ ratio versus ¹³C⁻/H⁻ ratio for Renazzo and Orgueil matrices and for the reference samples IOM-Gro, Mtm and IOM10. Each point corresponds to a 4 × 4 μm² ROI defined in NanoSIMS images. The D⁻/H⁻ ratio and ¹³C⁻/H⁻ ratio are linearly correlated for Orgueil and Murchison as well as in IOM10. (B) ¹⁶OD⁻/¹⁶OH⁻ ratio versus D⁻/H⁻ ratio for Renazzo and Orgueil matrices and for the reference samples IOM-Gro and Mtm (for clarity, only the mean values and 1σ standard deviations are shown, see Figure S2.A for details). The solid and dotted lines represent the regression line of pure components and its limits for a 95% prediction interval, respectively.

In Figure 3.B., the measured isotope data are shown in a plot of the ¹⁶OD⁻/¹⁶OH⁻ ratio as a function of D⁻/H⁻ ratio (details in Section 2.2). From the measured isotope ratios of Mtm, Ker and IOM, we determined the line of pure components and its associated 95% prediction interval (Figure 3.B). We used the data from the 4 × 4 μm² ROIs for Mtm (11 ROIs), Ker (8 ROIs) and IOM (16 ROIs); for clarity, these values are not shown in Figure 3.B. but in Figure S2.A. The data obtained for the Orgueil and Renazzo matrices systematically plot below the line of pure components. Because H⁻ intensity is higher for organic matter than for phyllosilicates, the former has a strong influence on the D⁻/H⁻ ratio of the mixture even when the organic/phyllosilicates ratio is low. Hence, a mixture of D-rich organics and D-poor phyllosilicates will follow a mixing curve that lies below the line of pure components.

According to Figure 3, carbonaceous chondrite matrices are characterized by a high concentration of hydrous minerals with low D/H ratios, mixed with a D-rich organic component. The quasi-linear correlation between the $D^-/H^-$ and $C^-/H^-$ ratios obtained for Orgueil and Renazzo, as well as for the synthetic mixture IOM10, can be used to estimate the isotopic ratio of the hydrous minerals with $C^-/H^- = 0$. For each sample, the intercept of the regression lines in the plot of $D^-/H^-$ versus $C^-/H^-$ was corrected for instrumental mass fractionation using the montmorillonite standard (Mtm) (Table 1). The D/H value obtained from the synthetic mixture IOM10 is slightly higher than the standard D/H ratio of the montmorillonite ($143 \times 10^{-6}$). Such isotopic enhancement may be the result of an imperfect linear emissivity of H in an organic/hydrous silicate mixture, resulting in an only partially linear correlation as indicated by linear correlation coefficients of 0.64 to 0.88 (Table 1). For the Orgueil and Renazzo chondrites, the estimated D/H ratios of H-bearing minerals are $155 \pm 14 \times 10^{-6}$ and $238 \pm 12 \times 10^{-6}$, respectively.

Table 1. Isotopic composition of the H-bearing minerals in IOM10, Orgueil and Renazzo matrix deduced from the $D^-/H^-$ versus $C^-/H^-$ plot: $D^-/H^-$ value for $C^-/H^- = 0$ ($D^-/H^-_{intercept}$) and D/H ratio corrected for instrumental mass fractionation (D/H $_{corrected}$ and associated δD). Error 1σ.

|  | $r_P$ [a] | $D^-/H^-_{intercept}$ ($\times 10^{-6}$) | D/H $_{corrected}$ ($\times 10^{-6}$) | δD $_{corrected}$ (‰) |
|---|---|---|---|---|
| IOM10 | 0.79 | 174 ± 15 | 162 ± 15 | 39 ± 97 |
| Orgueil | 0.64 | 166 ± 14 | 155 ± 14 | -7 ± 88 |
| Renazzo | 0.88 | 255 ± 9 | 238 ± 12 | 526 ± 88 |

[a] Linear correlation coefficient of Bravais-Pearson

3.2. Isotopic composition in the matrix of the Semarkona UOC

For the ordinary chondrite Semarkona, three pieces of matrix – Zone 1, Zone 5 and Zone 6 – were imaged with the NanoSIMS (Figure 2). Silicon covers large areas on the images while carbon is sparsely distributed as micrometer-scale spots in the $10 \times 10$ μm$^2$ areas.

The distributions of H$^-$ and $^{16}$OH$^-$ can be correlated visually but do not any show clear relationships with the distributions of $^{13}$C$^-$ or $^{28}$Si$^-$. The isotopic distributions of the three zones are very distinct (Figure 4). Zone 6 shows a quasi-homogeneous distribution of D$^-$/H$^-$ and $^{16}$OD$^-$/$^{16}$OH$^-$ ratios. Zones 1 and 5 contain large isotopic heterogeneities that are clearly visible in the $^{16}$OD$^-$/$^{16}$OH$^-$ ratio distribution. Zone 1 displays a bimodal distribution of D-poor pixels and pixels with $^{16}$OD$^-$/$^{16}$OH$^-$ ratios similar to those of Zone 6. This bimodal distribution is not visible for the D$^-$/H$^-$ ratios, but a wide peak is observed. Of note, a large number of pixels in each zone have D/H values within the same range as the homogeneous Zone 6 ($^{16}$OD$^-$/$^{16}$OH$^-$ ≈ $650 \times 10^{-6}$).

The Zone 1 images contain D-poor areas of 1 to 2 μm$^2$ in size (dark areas in the isotope distribution in Figure 2) and ~1 μm$^2$ sized D-rich areas (bright area in the bottom-left corner in Figure 2). In contrast, only D-rich areas are observed in Zone 5. Here, they present extreme D-enrichment compared to the surrounding matter and to the isotopic values of the other zones. Interestingly, D-rich areas are devoid of carbon in both Zones 1 and 5.

In contrast to the Orgueil and Renazzo carbonaceous chondrites, there is no clear relationship between D$^-$/H$^-$ and $^{13}$C$^-$/H$^-$ ratios in Semarkona (Figure 3 and 5). However, a large range of isotopic values was measured for the C-poor ROIs ($190 \times 10^{-6}$ < D$^-$/H$^-$ < $1950 \times 10^{-6}$), whereas a relatively low isotope ratio was measured for the C-rich ROIs (D$^-$/H$^-$ ≈ $350 \times 10^{-6}$). Note that the OD$^-$/OH$^-$ vs. $^{13}$C$^-$/OH$^-$ plot (not shown) shows identical features.

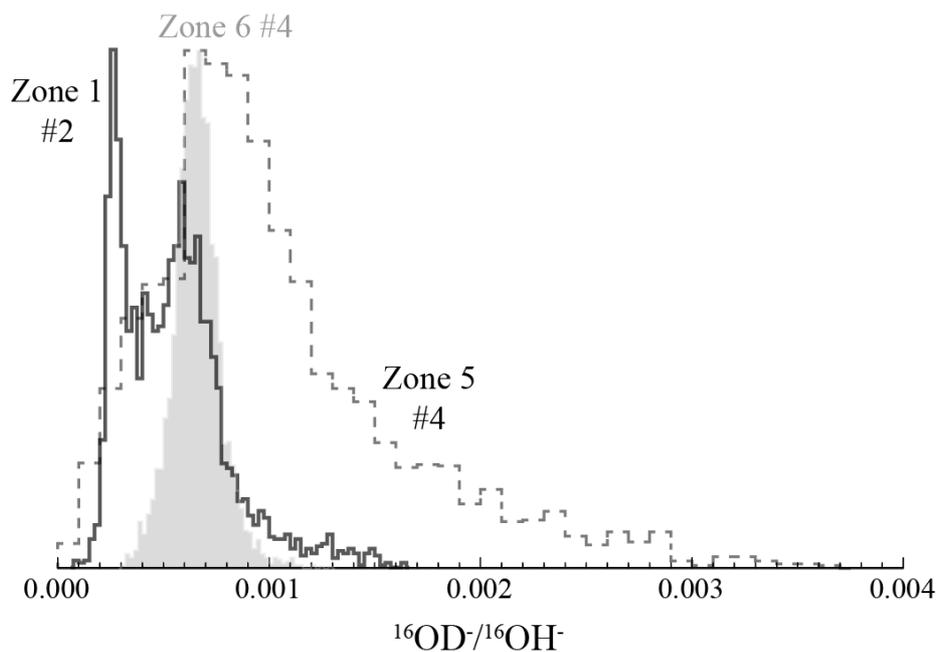

Figure 4. Pixel distribution of $^{16}OD^-/^{16}OH^-$ images in the three different matrix zones (Zones 1, 5 and 6) of Semarkona. The corresponding images are shown in Figure 2. Zone 6 (in grey) shows the narrowest distribution while Zone 1 (solid outline) and Zone 5 (dotted outline) contain D-poor and D-rich areas, respectively. The image number (#) of each sample is indicated for reference to the data in Table S2.

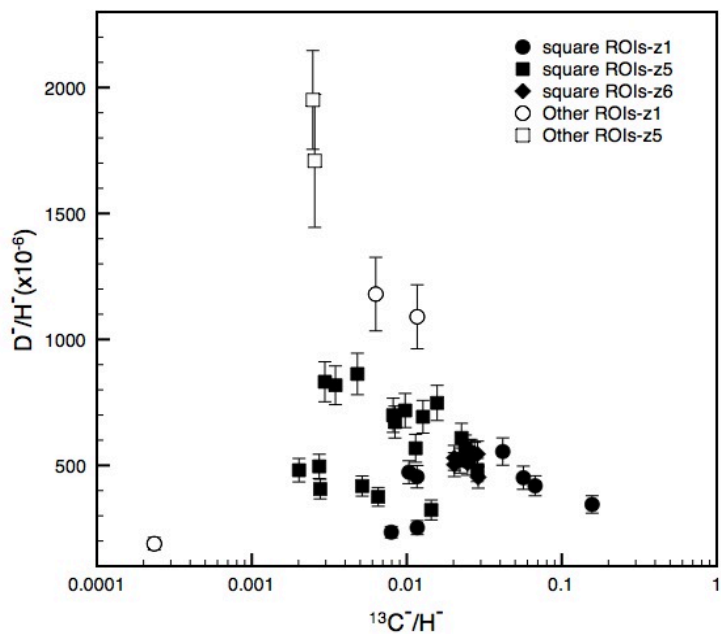

Figure 5. Measured $D^-/H^-$ ratio versus measured $^{13}C^-/H^-$ ratio for the three zones of the Semarkona matrix (circles = Zone 1, squares = Zone 5, diamonds = Zone 6). Each data-point corresponds to a 4 ×

4 μm² ROI (full symbols) or an ROI defined on the basis of isotope heterogeneities (open symbols) identified in the NanoSIMS images.

In Figure 6, measured isotope data are presented in a plot of the $^{16}OD^-/^{16}OH^-$ ratio vs. $D^-/H^-$ ratio for the three matrix zones analyzed in Semarkona. The line of pure components and its associated 95% prediction interval were determined using the data from the 4 × 4 μm² ROIs for Mtm (28 ROIs), Ker (12 ROIs) and IOM (24 ROIs); these values are presented in Figure S2.B. The majority of the matrix points plot on the line of pure components. However, 6 of the 32 square ROI data-points are located outside the prediction interval and above the line of pure components. All of these points belong to Zone 5 (filled squares in Figures 5 and 6). According to the model described in Section 2.2, these points can be interpreted as a mixture of D-poor organic matter and D-rich hydrous silicates. In the highly heterogeneous isotopic images (Figure 2), we isolated the most D-rich and the most D-poor regions for the three zones (represented by open symbols). Most of these correspond to C-poor areas with extremely high D-enrichments (Figure 6). As they plot on the line of pure components, we can assume that these ROIs correspond to quasi-pure hydrous silicates (i.e. with no organic H contribution). One area also shows a low C-content and a low D/H ratio (ROIs-z1) corresponding to D-poor phyllosilicates. However, the exact organic/hydrous silicate mixing-ratio cannot be qualitatively determined and this low D/H ratio should be considered as a maximum value for the hydrous silicates.

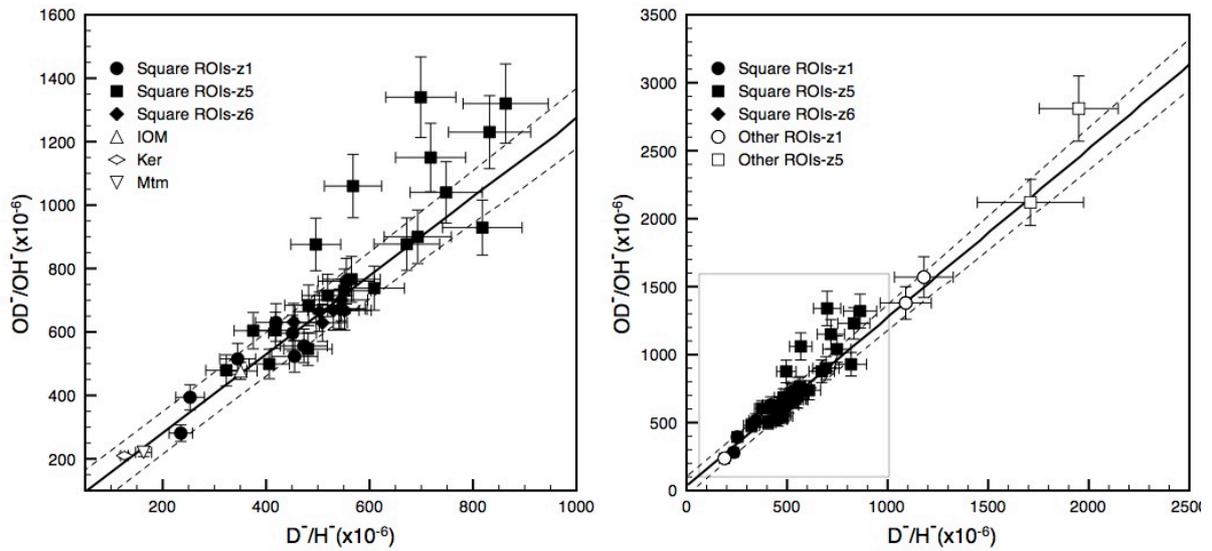

Figure 6. Measured $^{16}OD^-/^{16}OH^-$ ratio as a function of the measured $D^-/H^-$ ratio for the Semarkona matrix zones. Values measured for the reference samples IOM, Ker and Mtm are presented in the graph on the left. Values for the square ROIs (4 × 4 μm²) are shown as filled symbols (circles = Zone 1, squares = Zone 5, diamonds = Zone 6). The expanded graph on the right shows the values of the D-richest ROIs defined on the basis of image isotopic heterogeneity (open symbols). Error bars include internal and external precision at 1σ. The regression line of pure components (solid line) and the 95% prediction interval limits (dotted lines) are also shown.

We estimated the theoretical curve for a mixture of phyllosilicates with the highest D/H ratios in the isolated C-poor areas and organic matter with the D/H ratio of the C-rich areas (Figure 7). For this calculation, we used the mean value and standard deviation of IOM and Mtm intensities, as well as their isotopic fractionation ratios. Due to the difference in intensity of $H^-$ and $OH^-$ in phyllosilicates and organics, the mixing curve lies above the line of pure components. On the basis of this curve, the six data-points from Zone 5 that fall above the line of pure components can be easily explained as mixtures of a D-rich hydrous mineral and D-poor organic matter. The isolated ROIs reflect the extremely wide variation in the hydrogen isotopic composition of phyllosilicates in the matrix of the Semarkona chondrite. Moreover, many of the square ROI, which do not contain significant carbon, plot along the line of pure components. These can be interpreted as quasi-pure hydrous minerals with intermediate isotopic ratios.

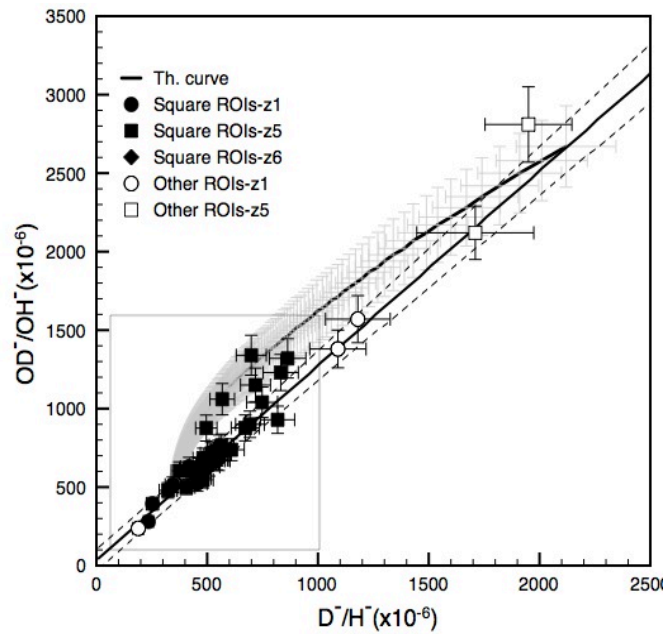

Figure 7. Theoretical mixing curve, calculated for a mixture of D-rich phyllosilicates and D-poor organic matter, superimposed on Figure 6. Error bars include internal and external precision at 1σ. The Zone 5 measurements that plot above the line of pure components can be explained by such mixing. The solid and dotted lines represent the regression line of pure components and the limits of the prediction interval at 95%, respectively. See Figure 6 for details of the data-points.

Measured hydrogen isotope ionic ratios were converted into absolute ratios for those isolated ROI identified as hydrous minerals, using terrestrial montmorillonite as a reference for the instrumental isotopic fractionation correction (Table 2). The corrected values reported in Table 2 were calculated using $^{16}OD^-/^{16}OH^-$ ratios to minimize the contribution of organic matter ($^{16}OH^-$ emission is higher in phyllosilicates than in organic matter). However, identical values are found when $D^-/H^-$ ratios are used.

To compare these five values with the mean hydrous silicate value measured in Semarkona, we also applied this instrumental isotopic fractionation correction to square areas characterized by a low C$^-$/H$^-$ ratio. We did not consider any ROI with C$^-$/H$^-$ values greater than 0.03, the mean value of the IOM10 mixture measured in the same session (Table S1).

Table 2. Isotopic composition of the hydrous silicates in Semarkona matrix: isolated ROIs and square ROI with C$^-$/H$^-$ < 0.03, and square ROI with C$^-$/H$^-$ < 0.03 in the homogenous Zone 6.

| ROI | D/H | er D/H | δD | er δD |
|---|---|---|---|---|
| | (×10$^{-6}$) | | ‰ | |
| Z5_1 | 1400 | 110 | 8,000 | 700 |
| Z5_2 | 1800 | 160 | 10,600 | 1030 |
| Z1_1 | 150 | 13 | -37 | 80 |
| Z1_2 | 900 | 80 | 4,800 | 510 |
| Z1_3 | 1020 | 100 | 5,600 | 640 |
| Sq. ROI[a] | 482 | 14[b] | 2100 | 90[b] |
| Z6[c] | 450 | 3[b] | 1900 | 20[b] |

[a] Sq. ROI = average value for square ROIs with C$^-$/H$^-$ < 0.03; [b] standard deviation of 44 ROIs; [c] Z6 = average value for square ROIs with C$^-$/H$^-$ < 0.03 in Zone 6 only.

The D/H ratios obtained for the five anomalous areas range from 150 × 10$^{-6}$ to 1800 × 10$^{-6}$. The latter value is the highest D/H ratio reported so far for hydrous minerals in the solar system. However, the extreme D/H variations we observed at the micrometer-scale are even more remarkable. The matrix surrounding each of these anomalous areas, composed mainly of an inorganic H-bearing component, has an intermediate D/H ratio close to 450 × 10$^{-6}$. Although both high and low D/H ratios are measured in the matrix, the average for all the 4 × 4 μm$^2$ ROIs (D/H = 480 × 10$^{-6}$) is close to this value, indicating that heterogeneity can also occur at a scale far lower than the spatial resolution of the NanoSIMS (<500 nm).

## 4. Discussion

4.1. Hydrothermal activity and organic matter/water interactions

4.1.1. Carbonaceous chondrites

Carbonaceous chondrite matrices are characterized by a high concentration of a mineral H-bearing component with low D/H ratios, mixed with a D-rich organic component. The correlation between $D^-/H^-$ and $C^-/H^-$ ratios argues for a simple mixing between two phases with a relatively well-defined D/H ratio. The estimated D/H ratios for phyllosilicates in the matrices of Orgueil and Renazzo are $155 \pm 14 \times 10^{-6}$ and $238 \pm 12 \times 10^{-6}$, respectively.

Our results are consistent with D/H ratios reported in previous studies. The D/H ratio of H-bearing silicates can be calculated using D/H ratios measured for isolated organics and bulk chondrite organic matter content. IOM in Orgueil has been measured after acid demineralization and is estimated to have a D/H ratio of $327 \times 10^{-6}$ in Yang and Epstein (1983) and $307 \times 10^{-6}$ in Alexander et al. (2007). In Renazzo, acid-isolated organic matter is apparently D-richer, with a bulk D/H value of $483 \times 10^{-6}$ (Yang and Epstein, 1983). However, both CR and CI IOM, which occur as separate grains in the matrix (Remusat et al., 2010; Le Guillou et al., 2014), are heterogeneous with respect to their D/H values when imaged at the micrometer-scale by NanoSIMS (Busemann et al., 2006; Remusat et al., 2009). The D/H ratio of phyllosilicates in Orgueil has been estimated from mass-balance calculations to be quasi-terrestrial (Robert and Epstein, 1982; Alexander al., 2010) or even lower, with D/H from 64 to $98 \times 10^{-6}$ (Alexander et al., 2012). Renazzo phyllosilicates are believed to have heavier compositions (D/H = $262 \times 10^{-6}$, Robert et Epstein, 1982) than Orgueil phyllosilicates. In situ measurements have shown that CR phyllosilicates may have a D/H ratio above $320 \times 10^{-6}$ (Deloule and Robert, 1995) or in the range of 180 to $220 \times 10^{-6}$ (Bonal et al., 2013).

Recently, Alexander et al. (2012) estimated the D/H ratio of the initial water accreted on the carbonaceous chondrite parent body using the bulk isotopic and elemental compositions measured for a set of CR, CM and CI chondrites. In a plot of δD vs. C/H, they found that the different CR and CM-CI chondrites define quasi-parallel lines showing a linear increase of δD with the C/H ratio for each group. The intercept of the line at C/H = 0 gives the water D/H ratio ($[D/H]_{C/H=0}$) in CM-CI and CR chondrites: $[D/H]_{C/H=0}$ = 86.6 ± 3.6 × $10^{-6}$ and $171^{+17}_{-10}$ × $10^{-6}$, respectively. Our results confirm these correlations but we obtained different values for $[D/H]_{C/H=0}$ (see Table 1); 155 ± 14 ×$10^{-6}$ for Orgueil and 238 ± 12 ×$10^{-6}$ for Renazzo. The difference between the results of the two studies should, however, be taken with caution as variations in a micrometric region of a single sample are being compared with variations between bulk chondrites belonging to the same family but with a substantial difference in their alteration degree.

The correlation between bulk D/H and C/H ratios for CM and CI chondrites and CR chondrites is easily explained by a mixing of different proportions of two major H-bearing components: D-poor water preserved in hydrous minerals and D-rich organic matter (Alexander et al. 2010). The preservation of these correlations requires that aqueous alteration and metamorphism occurred in a closed-system with respect to hydrogen. Such a hypothesis is consistent with petrographic and geochemical observations in carbonaceous chondrites (Bland et al., 2009). These authors have re-estimated the permeability of primitive carbonaceous chondrite to be lower that previously assumed ($10^{-17}$ to $10^{-19}$ $m^2$). This low permeability results in very limited water flow and in aqueous alteration that extends over distances of a few hundred μm at most.

Taking into account the correlation between D/H and C/H ratios that we observed within the Orgueil and Renazzo matrices, it seems that no equilibration occurred between water and organics, even at the scale of a few micrometers. This is consistent with a previous in situ study (Remusat et al., 2010). The correlated variations at a micron scale between D/H and C/H would thus be caused by the relative proportions of IOM and phyllosilicates sampled by the ion beam, as observed for the IOM10 mixture.

Further analysis of the D/H and C/H ratios in Orgueil, Renazzo and other CI, CM and CR chondrites are required to determine whether the signature of initial water has been preserved in the primitive chondrites.

4.1.2. Unequilibrated ordinary chondrite

In contrast to the carbonaceous chondrites, the matrix of the Semarkona ordinary chondrite contains H-bearing silicates exhibiting a wide range of D/H ratios, mixed with a relatively D-poor organic component (Figure 5). Hydrous minerals are highly heterogeneous with respect to their D/H ratios, even at the micron scale; in a single image, pixel distribution can vary by up to one order of magnitude (Figure 4) and a bimodal distribution is observed in Zone 1. Moreover, the three pieces of matrix that we analyzed exhibit quite different D/H distributions, attesting to the highly heterogeneous character of the matrix at the millimeter to centimeter scale. Nevertheless, a common D/H value of ~ $450 \times 10^{-6}$ is obtained for a representative number of pixels in the isotopic images of each zone.

Even though previous estimates for the bulk Semarkona organic matter differ – a mean D/H ratio of $263 \times 10^{-6}$ was reported by Yang and Epstein, (1983), while Alexander et al. (2007) measured a ratio of $517 \times 10^{-6}$ – this difference is relatively small compared to the range that we observed for hydrous minerals. Some local variations within areas of a few tens of micrometers in size have also been observed for Semarkona IOM by NanoSIMS imaging (Remusat and Piani, 2013), but these remain within the range defined here for organic matter. In situ ion-probe measurements of phyllosilicates in the Semarkona matrix have reported high D/H values, with variations from 670 to $870 \times 10^{-6}$ determined with a 10 μm diameter primary beam (Deloule and Robert, 1995). Taking into account the differences in spatial resolution (~500 nm in our study), our results are consistent with these results and considerably widen the range of variation in D/H values so far reported for H-bearing silicates (D/H higher than $1800 \times 10^{-6}$).

It is pertinent to highlight here the issue of standard calibration. Because the hydrogen isotopic ratios measured for Semarkona hydrous silicates far exceed the D/H ratios of terrestrial objects, no D-rich phyllosilicates with a known D/H ratio calibrated relative to the SMOW are currently available for use as a reference sample. As in previous studies (e.g. Deloule and Robert, 1995, 1998, Bonal et al. 2013), we extrapolated the calibration line derived from a terrestrial reference sample towards high D/H values. But no precise investigation has yet been carried out into the possible bias introduced by this extrapolation. A D-rich standard could also be used to precisely quantify any terrestrial contamination. Due to its crystalline nature, smectite can accommodate variable amounts of water between its silicate layers. This interlayer water is known to exchange rapidly with atmospheric water (Savin and Epstein, 1970). As a consequence, our corrected D/H ratios for hydrous silicates are probably underestimated and should be considered minimum values. Despite this issue, our results provide new information concerning the spatial distribution of the hydrous silicate phase and its water precursor at the micron scale.

It has been proposed that the D-enrichments found in OC could have resulted from the reduction of water induced by metal oxidation during aqueous alteration and metamorphism, which would have led to a Rayleigh distillation process (Alexander et al., 2010). Isotopic exchange between water and organics would have then occurred, producing the D-enrichment found in the IOM of unequilibrated OC of higher petrographic degree (Alexander et al., 2010). This process would have required a large amount of initial water because 95% to more than 99% of the water would have needed to have been converted into $H_2$ to have induced the large D-enrichments observed. In addition, to account for the occurrence of this process, a significant amount of iron must have been oxidized in the extremely D-rich areas that we observed.

We thus favor the more simple assumption that the hydrogen isotopic heterogeneity was inherited from the ice precursors. This would have resulted in the accretion of water with a variable D/H ratio, and the extreme D/H values we observed at the micron-scale therefore argue for a low degree of hydrothermal activity in the Semarkona parent-body. Petrographic observations have shown that hydrous minerals in primitive LL3 ordinary chondrites are mainly found in the fine-grained and more reactive matrix (Alexander et al., 1989; Hutchison et al., 1987). However, our study provides new limits for this alteration, showing that, even in the fine-grained matrix, isotopic homogenization has not been achieved at the micrometer scale. Moreover, the wide variation in the phyllosilicate D/H ratio coupled with the relatively homogeneous and intermediate isotopic composition of organic matter argues against equilibrium exchange between these two phases.

Observation of such a heterogeneous D/H composition in the Semarkona matrix raises the question of the conditions of formation of the phyllosilicates. Considering that hydrogen isotope fractionation between phyllosilicates and water is negligible (just a few percent, e.g. Savin and Epstein, 1970) compared to the range of variation measured in the Semarkona matrix, our observations imply that the water from which the Semarkona phyllosilicates were formed was isotopically heterogeneous. Phyllosilicates were formed in the parent body through localized water-rock interactions in the matrix. In this scenario, the water that accreted as ice grains was isotopically heterogeneous and has transferred its heterogeneity to the phyllosilicates. The size of the reaction zone was limited to a few micrometers, as observed by the NanoSIMS. The phyllosilicate precursors could have been incorporated into the matrix along with ice grains (at T<200K) in the form of amorphous silicates, condensed from a vapor, and known to be highly hygroscopic (Le Guillou et Brearley, 2013).

## 4.2. Pristine D-rich water reservoir recorded in ordinary chondrites: implication for water in the early solar system

Some hydrous minerals in Semarkona have recorded the signature of heavy water. This water had D/H values of at least $1800 \times 10^{-6}$, attesting to the existence of a D-rich reservoir of water in the early solar system. Moreover, we observed that the D-rich silicates in Semarkona are surrounded by silicates with moderate D/H ratios ($\sim 450 \times 10^{-6}$) and we measured even lower values in some hydrous minerals (D/H ratios down to $\sim 150 \times 10^{-6}$) (Table 2).

The hydrogen isotopic composition of water recorded in Semarkona is unusual in its heterogeneity as well as in its extreme D-enrichment compared to other objects measured so far in the solar system (Figure 8). Indeed, a narrow range of hydrogen isotopic compositions is found for water in around one hundred carbonaceous chondrites, on Earth, for the Jupiter family comet Hartley 2 and for the Semarkona water D-poor end-member. Only some Oort cloud comets and Enceladus water display significantly higher D/H ratios ($\sim 300 \times 10^{-6}$), but this value remains relatively low compared to the wide range found for LL chondrite water.

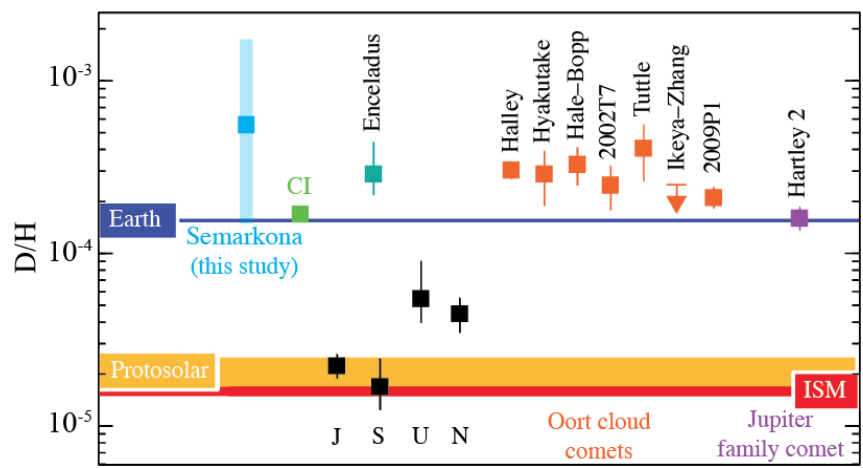

Figure 8. D/H ratios of Semarkona hydrous minerals compared to carbonaceous meteorites (CI), the Earth's oceans, Enceladus and cometary water. For Semarkona, the blue square is the mean D/H ratio

of Zone 6 and the thick line shows the range of D/H values recorded; Table 2. Data for planets, interstellar medium (ISM), and the protosolar nebula refer to the value in $H_2$. Adapted from Hartogh et al. (2011) and Bockelée-Morvan et al., (2012).

In the interstellar medium, a large range of D/H values has been observed for water in protostar environments (~$10^{-4}$ to $10^{-2}$, Ceccarelli et al., 2014). However, an upper limit of ~ 5 × $10^{-3}$ has been estimated for the D/H ratio of protostar water ice (Dartois et al., 2003). Based on previous calculations, the D/H ratio of water is expected to be low in the inner part of the solar disk, due to intensive exchange with solar $H_2$ gas (D/H = 20 × $10^{-6}$), and higher in the outer part (see Section 1.). In their model, Yang et al. (2013) considered the dynamic supply of D-rich water from the molecular cloud during the evolution of the early solar system (first million years). This model predicts the formation of a D-rich plateau at around 10 AU, propagating towards the outer disk with time, and allows similar D/H ratios for water formed at different heliocentric distances to be obtained.

Because Semarkona displays D/H ratios that are significantly higher than the chondritic value (D/H = 150 × $10^{-6}$), and as the lowest D/H ratios measured for hydrous silicates in Semarkona are chondritic, one might consider that this meteorite accreted both the homogenized chondritic water and a D-rich water ice similar to that that we measured in the matrix (D/H = 1800 × $10^{-6}$). Given that Oort-cloud family comets show significant deuterium enrichment relative to chondritic water, we can postulate that Oort-cloud family comet water is also a mixture of chondritic water and D-rich water. Their water D/H ratios can therefore be used to estimate the relative proportions of these two components. In the Semarkona chondrite, the homogeneous Zone 6 exhibits the mean D/H ratio of water (D/H = 450 × $10^{-6}$). Thus, we can calculate that 82% would be chondritic and 18 % would be derived from D-rich water ice. For a cometary D/H ratio of $300 \times 10^{-6}$, 91% would originate from chondritic water and 9 % from D-rich pristine solar system water.

In a recent model, ion-molecule reactions in the conditions of the protoplanetary disk were proposed to be poor at enhancing the D/H ratio of water (Cleeves et al., 2014). According to Cleeves et al., at least 7% of the chondritic-like water is inherited from the interstellar medium (D/H ~ $2 \times 10^{-3}$). Using the highest D/H ratio that we measured in Semarkona ($1.8 \times 10^{-3}$), a similar value is obtained (8%). Because the chondritic-like water D/H ratio seems to have been widely distributed among the inner solar system objects (Figure 8), it seems that the proportion estimated by Cleeves et al. (2014) is representative of the main reservoir of water in the inner solar system.

However, we further suggest that LL chondrites and Oort-cloud family comets could be the results of an extra-addition of interstellar-like water (D/H ~ $2 \times 10^{-3}$) to this chondritic-like water (D/H ~ $150 \times 10^{-6}$), and that the end-members of this subsequent mixing are partly preserved in the matrix of Semarkona. This hypothesis has two main implications: (i) that the Oort-cloud family comets and the LL chondrites formed after the main period of water homogenization in the inner solar system and (ii) that a part of the interstellar D-rich ice remained preserved at the times and places where LL ordinary chondrites and Oort-cloud family comets formed.

## 5. Conclusion

We performed NanoSIMS hydrogen isotopic measurements in the matrices of two carbonaceous chondrites (Orgueil and Renazzo) and the Semarkona unequilibrated ordinary chondrite. Using the protocol developed in Piani et al. (2012), we analyzed areas of matrix that consist of intimate mixtures of organics and hydrous minerals.

In carbonaceous chondrite matrices, D-rich organic matter is mixed with homogeneous D-poor H-bearing minerals, while the opposite is observed in the matrix of Semarkona. Hydrous minerals in the Semarkona matrix display a wide range of D/H compositions (from 150 up to 1800 × $10^{-6}$). These isotopic variations occur at the micrometer scale. These values are considerably higher than that found for Semarkona organic matter and demonstrate that: (1) exchange with organics was not the cause of hydrous silicate isotopic heterogeneity; and (2) the extent of aqueous alteration is limited to a few microns. Phyllosilicate isotopic heterogeneity must therefore be the consequence of the accretion of water-ice precursors that display a large range of isotopic compositions.

A D-rich reservoir was partly preserved in the LL ordinary chondrite parent body and could also be the source of part of the water outgassed from the Oort cloud family comets. Understanding the origin of the relative solar system homogeneity and why ordinary chondrites and Oort cloud family comets were able to preserve some heterogeneity would constitute a significant step towards unraveling the origin of solar system water.

**Acknowledgments**

Jean Duprat, Etienne Deloule, Yves Marrocchi, Corentin Le Guillou and Shoichi Itoh are thanked for stimulating discussions. Larry Nittler and one anonymous reviewer are thanked for their thorough and constructive comments. We are also grateful to Editor Tim Elliott for his helpful reviews and careful editing. This work has been supported by the Programme National de Planétologie (PNP INSU) and the ANR T-Tauri Chem. The National NanoSIMS facility at the Muséum National d'Histoire Naturelle was established by funds from the CNRS, Région Ile de France, Ministère délégué à l'Enseignement supérieur et à la Recherche, and the Muséum itself.